\documentstyle[12pt]{article}
\input{epsf.tex}
\begin{document}

  \begin{flushright} \begin{small}
  CERN-TH/99-402 \\DF/IST-8.2000 \\hep-th/0011234
  \end{small} \end{flushright}
\vskip1cm

\begin{center}
{\Huge\bf Supersymmetry of the extreme rotating toroidal black hole}  

\vskip 1cm

 {\bf Jos\'e P. S. Lemos}\footnote{E-mail: lemos@kelvin.ist.utl.pt }

\smallskip
  { CENTRA, Departamento de F\'{\i}sica,
              Instituto Superior T\'ecnico,} \\
  { Av. Rovisco Pais 1, 1096 Lisboa, Portugal,} 
\\ \& \\
{ Theory Divison, CERN, CH-1211, Geneva 23, Switzerland.} \\

\end{center}
\date{\today}
\vskip1cm
\begin{abstract}
We study the supersymmetry of the charged rotating toroidal black hole
solutions found by Lemos and Zanchin, and show that the only
configurations that are supersymmetric are: (i) the non-rotating
electrically charged naked singularities already studied by Caldarelli
and Klemm, and (ii) an extreme rotating toroidal black hole with zero
magnetic and electric charges.  For this latter case, the extreme
uncharged black hole, we calculate the Killing spinors and show that
the configuration preserves the same supersymmetries as the background
spacetime.
\end{abstract}
\newpage

\noindent
{\bf 1. Introduction}
\vskip 3mm

There are several motivations to, first, find supersymmetric black hole
solutions and, then, analyze their properties. For instance, black
holes that are supersymmetric, have zero temperature \cite{gibbons1},
and in some cases it is possible to associate them with solitons of the
theory, interpolating between two distinct homogeneous vacua solutions
\cite{gibbons2}.  In addition, in certain instances, supersymmetric
black holes are exact solutions of the theory even when quantum
corrections are taken into account \cite{kalosh92}.

Previous studies looking for supersymmetric configurations appeared in
the context of $N=2$ ungauged supergravity whose bosonic sector is pure
Einstein-Maxwell theory and for which the black hole solutions, the
Kerr-Newman solutions, are asymptotically flat.  It was found that the
Kerr-Newman family is supersymmetric, i.e., admits Killing spinors,
when the mass of the solution is equal to the charge. Thus for non-zero
angular momentum the supersymmetric solutions are naked singularities,
whereas for zero angular momentum, the supersymmetric solution is the
extreme Reissner-Nordstr\"om black hole \cite{gibbons3,todd}.

Subsequent studies appeared within the context  of $N=2$ gauged
supergravity, obtained by gauging the symmetry that rotates the fermions 
(gravitini) of the ungauged theory, in which case the bosonic
sector is Einstein-Maxwell with a negative cosmological constant.  Due
to the presence of the negative cosmological constant there are now
three classes of black hole solutions in which the horizons can have
spherical, toroidal (cylindrical and planar), and hyperbolical
topologies. The supersymmetry of the spherical solutions studied in
\cite{romans92,kosteleckyperry96} show that to be supersymmetric the
black hole must rotate. Non-rotating supersymmetric configurations are
naked singularities. The supersymmetries of toroidal and hyperbolic
topologies were studied in a comprehensive paper by Caldarelli and Klemm
\cite{caldarelliklemm98}. For toroidal black holes a result similar to
spherical black holes was obtained, while for hyperbolical black holes
there are also non-rotating configurations, with magnetic charge, which
are supersymmetric \cite{caldarelliklemm98}.  The study of
supersymmetries were extended in \cite{alonsoalercaetal00} to include
Taub-NUT spacetimes, all of these solutions belonging to the
Plebanski-Demianski family of solutions.

A different rotating toroidal (or cylindrical) black hole solution was
found in \cite{lemos1} and then generalized to include charge
\cite{lemoszanchin96}, (see also \cite{lemos2}, \cite{lemosreview} for
a review, and the charged solution in the non-rotating case is also
discussed in \cite{huang}).  These solutions are not isometric to the
toroidal solutions found by Klemm and Vanzo \cite{klemm}, which are a
special case of the Plebanski-Demianski metrics, and whose
supersymmetries were studied in \cite{caldarelliklemm98}.  The main
aim of this paper is to study the supersymmetries of the toroidal (or
cylindrical) rotating charged black holes in anti-de Sitter spacetime
found in \cite{lemos1,lemoszanchin96}.  We study the integrability
conditions, and find that the supersymmetric black hole has zero
charge, must rotate with extreme angular velocity and thus has zero
temperature.  The corresponding Killing spinor is explicitly displayed
and it is shown that the extreme black hole preserves the same number
of supersymmetries as the original anti-de Sitter spacetime.  There
are other non-rotating solutions, representing charged naked
singularities, which are supersymmetric \cite{caldarelliklemm98}.

\vskip 0.5cm

\noindent
{\bf 2. $N=2$ gauged supergravity}

\vskip 3mm

In $N=2$ supergravity there are four bosonic degrees of freedom,
represented by a graviton $e_m^a$ and a Maxwell gauge field $A_m$, and
four fermionic degrees of freedom, represented by two real gravitini
$\psi_m^i$ $(i=1,2)$, usually combined into a single complex spinor
$\psi_m = \psi_m^1 + i\psi_m^2$.  Gauged $N=2$ supergravity can be
obtained from the ungauged theory by gauging the $\mathrm{SO(2)}$
symmetry that rotates the gravitini, with the corresponding
appearance of a minimal gauge coupling $\alpha$ between the photons and the
gravitini. Invariance of the action under local supersymmetric
transformations requires a mass term for the gravitini and the
introduction of a negative cosmological constant.

The action is 
\begin{eqnarray}
S &=& \int d^4x \,\, \lbrace
            R + 6\alpha^2 -F_{mn}F^{mn} -2 \bar{\psi}_m\gamma^{mnp}D_n\psi_p
            + 2\alpha \bar{\psi}_m\gamma^{mn}\psi_n \nonumber
                            \\
          &&- \frac{i}{2} (F^{mn} + \hat{F}^{mn})
            \bar{\psi}_p\gamma_{\left[m\right.}\gamma^{pq}
             \gamma_{\left.n\right]}
            \psi_q     \rbrace.
            \label{action1}
\end{eqnarray}
We use the same notation as in \cite{caldarelliklemm98}:  
$a,b,\ldots$ are tangent space
indices, and $m,n,\ldots$ are world indices. The signature
is $(-,+,+,+)$, and the real representation of the
gamma matrices $\gamma_a$ is used.  The $\gamma$ matrices satisfy 
$\{\gamma_a,\gamma_b\}=2\eta_{ab}$, antisymetrization 
is defined by $\gamma_{ab} \equiv \gamma_{\left[a\right.}
\gamma_{\left.b\right]} \equiv \frac{1}{2}[\gamma_a,\gamma_b]$, and 
$\gamma_5 = \gamma_{0123} = \gamma_0\gamma_1\gamma_2\gamma_3$.  From
equation (\ref{action1}) one sees that the cosmological constant
$\Lambda$ is related to the gauge coupling $\alpha$ by $\Lambda =
-3\alpha^2$.  $D_m$ is the gauge covariant derivative given by
\begin{equation}
D_m = \nabla_m - i \alpha A_m, \label{derivative1}
\end{equation}
with $\nabla_m$ being  the Lorentz covariant derivative
\begin{equation}
\nabla_m = \partial_m + \frac{1}{4}\omega_m^{\;\;ab}\gamma_{ab}.
                                               \label{derivative2}
\end{equation}
The spin connection  is given in terms of the 
tetrads and spinors by 
\begin{eqnarray}
\omega_{mab} &=& 
              \;\;\;e^n_a \lbrace\partial_{\left[m\right.}e_{\left.bn\right]}
              - \frac{1}{2}{\mathrm{Re}}(\bar{\psi}_m\gamma_b\psi_n) \rbrace
                      \nonumber \\
             &&-e^n_b \lbrace\partial_{\left[m\right.}e_{\left.an\right]}
                - \frac{1}{2}{\mathrm{Re}}(\bar{\psi}_m\gamma_a\psi_n) \rbrace
                  \nonumber \\
  &&-e_a^l e^n_b e_m^c \lbrace\partial_{\left[l\right.}e_{\left.cn\right]}
    - \frac{1}{2}{\mathrm{Re}}(\bar{\psi}_l\gamma_c\psi_n) \rbrace\,.
                                                     \label{spinconnection1}
\end{eqnarray}
$F_{mn}$ and $\hat{F}_{mn}$ denote the standard and the supercovariant 
field strength, respectively, with
\begin{equation}
\hat{F}_{mn} = F_{mn} - {\mathrm{Im}}(\bar{\psi}_m\psi_n) \, . 
                                                 \label{supermaxwell}
\end{equation}
The action (\ref{action1}) is invariant under the usual general coordinate, 
local Lorentz, and $\mathrm{U(1)}$ gauge transformation. It is also invariant 
under the following local $N=2$ supersymmetric transformations
\begin{equation}
\delta e_m^{a} = {\mathrm{Re}}(\bar{\epsilon}\gamma^a\psi_m), \quad
\delta A_m = {\mathrm{Im}}(\bar{\epsilon}\psi_m), \quad
\delta \psi_m = \hat{\nabla}_m \epsilon\, ,
                             \label{supersymmetrytrans}
\end{equation}
where $\hat{\nabla}_m$ is the supercovariant derivative defined by
\begin{equation}
\hat{\nabla}_m = D_m + \frac{1}{2\ell}\gamma_m + \frac{i}{4}
\hat{F}_{ab}\gamma^{ab} \gamma_m \, ,       
                                      \label{derivative3}
\end{equation}
and $\epsilon$ is an infinitesimal Dirac spinor. 

To study a theory one should look for solutions of the equations of
motion.  One way to find solutions of action (\ref{action1}) is to
consider its bosonic sector alone, by putting the fermionic fields equal
to zero, $\psi_m=0$, and then find solutions within this sector. A
solution of the bosonic sector is also a solution of the full
supergravity theory if the solution admits a Dirac Killing spinor
$\epsilon$ whose supercovariant derivative is zero, $\hat{\nabla}_m
\epsilon=0$. Indeed, in this case, the last equation of
(\ref{supersymmetrytrans}) defines a Killing spinor equation for the
Killing spinor $\epsilon$, and tells that the fermionic fields remain
identically zero for any supersymmetric transformation. The
corresponding bosonic solution is then a true solution of the theory.

Thus, among the solutions of the bosonic sector one looks for 
those which are invariant under supersymmetric transformations, 
i.e., solutions which admit a Killing spinor obeying the 
equation
\begin{equation}
\hat{\nabla}_m \epsilon = 0. 
                              \label{killingspinorequation}
\end{equation}
The integrability conditions for (\ref{killingspinorequation}) are then
\begin{equation}
\hat{R}_{mn}\epsilon \equiv \left[\hat{\nabla}_m,\hat{\nabla}_n\right]
\epsilon =0, 
                              \label{integrabilityconditions}
\end{equation}
where $\hat{R}_{mn}$ is the supercurvature. Note that since
(\ref{integrabilityconditions}) is a local equation,  it gives a
necessary condition for the existence of Killing spinors, but not a
sufficient one.  In order to assure that there are Killing spinors one
has to integrate equation (\ref{killingspinorequation}).  In the
following sections, we shall solve (\ref{killingspinorequation}) and
(\ref{integrabilityconditions}) for the black hole spacetimes
introduced in \cite{lemos1,lemoszanchin96}.

\vskip 0.5cm

\noindent
{\bf 3. The toroidal solutions in general relativity}

\vskip 3mm

By putting to zero the fermionic sector of $N=2$ supergravity 
in (\ref{action1}) one obtains the Einstein-Maxwell action  
\begin{equation}
S = \int d^4x \left( R + 6\alpha^2
            - F_{mn}F^{mn}\right)   .
            \label{action2}
\end{equation}
A rotating solution, which can represent a black hole, may be found
from the equations of motion following from (\ref{action2}) (see
\cite{lemoszanchin96}).  The rotating solution can be obtained from the
non-rotating toroidal metric by mixing time and angle into a new time
and a new angle. Since angles are periodic, this is not a proper
coordinate transformation, yielding in this way a new solution globally
different from the static one (see \cite{lemos1}).  This metric is thus
different  from the rotating toroidal metric found by Klemm and Vanzo
\cite{klemm} which cannot be obtained by this forbidden
coordinate mixing, is instead obtainable from the general Petrov
type-$D$ solution. Our charged rotating toroidal solution is
given by
\begin{eqnarray}
ds^{2} &=& - \frac{\Delta}{(\alpha r)^2} (dt-\lambda d\phi)^2 + 
\frac{(\alpha r)^2}{\Delta} dr^2 +
\nonumber\\
&&+(\alpha r)^2 dz^2 + r^2(\alpha^2 \lambda dt - d\phi)^2\, ,
                                                       \label{metric1}
\end{eqnarray}
\begin{equation}
A= -\frac{2Q_E}{\alpha r} (dt-\lambda d\phi) - 
    2Q_M z (d\phi-\lambda \alpha^2 dt)\, .
                                                      \label{potential1}
\end{equation}
The Ricci curvature $R_{mn}$ can be found from (\ref{metric1}), which
we will do later. The Maxwell tensor $F_{mn}$ can be found from the
Maxwell curvature 2-form $F=F_{mn}dx^m\wedge dx^n$. From
(\ref{potential1}) it is given by
\begin{equation}
F= dA= \frac{2Q_E}{\alpha r^2} dr \wedge (dt-\lambda d\phi) 
      -2 Q_M dz \wedge (d\phi-\lambda \alpha^2 dt) \, .
                                                           \label{maxwell1}
\end{equation}
In this solution, the ranges of the time and radial coordinates are
$-\infty<{t}<+\infty$, $0\leq r< +\infty$.  
The topology of the two dimensional space, $t={\rm constant}$ and 
$r={\rm constant}$, generated by the group $G_2$ can be (i)
$R\times S^1$, the standard cylindrical symmetric model, with orbits
diffeomorphic either to cylinders or to $R$ (i.e, $G_2=R\times U(1)$),
with  $-\infty <z<+\infty$, $0\leq{\phi}< 2\pi$, 
(ii) $S^1 \times S^1$ the flat torus $T^2$ model ($G_2=U(1) \times
U(1)$) with  $0\leq \alpha z< 2\pi$, $0\leq{\phi}< 2\pi$, and (iii)
$R^2$, the planar model with $-\infty <z<+\infty$,
$-\infty <(\phi/\alpha)<+\infty$, but in this case the black hole
(in fact, a black membrane) does not rotate.

In order to proceed, we choose the following tetrads 
\begin{equation}
e^0= \frac{\sqrt\Delta}{\alpha r}(dt -\lambda d\phi)\, , 
\quad e^1=\frac{\alpha r}{\sqrt\Delta} dz \, ,
\quad e^2=\alpha r dz\, ,
\quad e^3=r(\lambda \alpha^2 dt - d\phi) \, .
                                                           \label{tetrads1}
\end{equation}
Then  equations (\ref{metric1})-(\ref{potential1}) can be written 
as
\begin{equation}
ds^2=-(e^0)^2+(e^1)^2+(e^2)^2+(e^3)^2
\, ,
                                                           \label{metric2}
\end{equation}
\begin{equation}
A=-\frac{2Q_E}{\Delta}e^0 + 2Q_M\frac{z}{r} e^3 \, .
                                                           \label{potential2}
\end{equation}
The Maxwell 2-form is 
\begin{equation}
F= -\frac{2Q_E}{\alpha r^2}e^0\wedge e^1 
+ \frac{2Q_M}{\alpha r^2} e^2\wedge e^3
\, .
                                                           \label{maxwell2}
\end{equation}
The functions and parameters appearing in
(\ref{metric1})-(\ref{maxwell2}) are:  $\Delta= 4c^2 -b \alpha r +
\alpha^4 r^4$, $\lambda=\frac{1}{\alpha}
\frac{\sqrt{1-\sqrt{1-\frac{8J^2\alpha^2}{9M^2}}}}
{\sqrt{1+\sqrt{1-\frac{8J^2\alpha^2}{9M^2}}}}$ a specific angular
momentum parameter, $J$ and $M$ being the angular momentum and mass per
unit length of the system, respectively,
$c^2=4Q^2(1-\lambda^2\alpha^2)$ with $Q^2=Q_E^2+Q_M^2$, $Q_E$ and $Q_M$
being the electric and magnetic charges per unit length, respectively,
and $b=4M\frac{1-\lambda^2\alpha^2}{1+\frac12 \lambda^2\alpha^2}$.
Note that $M=0$ implies $J=0$.  In \cite{lemoszanchin96} another
specific angular momentum parameter $a$ was used, related to $\lambda$
by $\lambda=a/\sqrt{1-\frac12a^2\alpha^2}$.  For the toroidal case, the
mass $\bar M$, angular momentum $\bar J$, and charge $\bar Q$ are given
in terms of $M$, $J$ and $Q$ by $\bar{M}=\frac{2\pi}{\alpha} {M}$,
$\bar{J}=\frac{2\pi}{\alpha} J$, and $\bar{Q}=\frac{2\pi}{\alpha} Q$
\cite{lemoszanchin96}.  When $\Delta$ has two roots the solution
represents a black string (in the cylindrical model) or a black hole
(in the toroidal model), when it has one root it represents,
correspondingly, an extreme black string or an extreme black hole, and
when it has no roots it represents, correspondingly, a singular naked
straight line or a singular closed line.
\vskip 1mm
\centerline{\epsffile{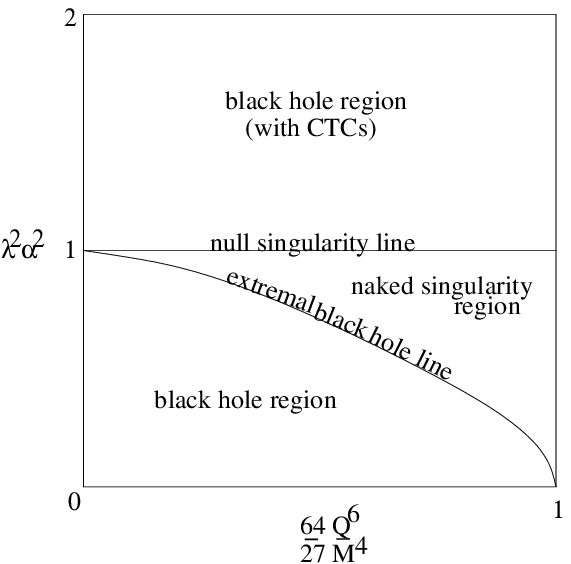}} 
{\noindent Figure 1. The five regions and lines which 
yield solutions of different nature are shown in the parameter 
space $\lambda\times Q$ in appropriate units of $\alpha$ and mass, 
respectively. }
\vskip 3mm

In figure  1, we show the black hole and naked singularity regions, 
and the extremal black hole line dividing those two regions, as well as 
the other solutions in the upper part of the figure. The regions are
\begin{description}
\item[(i)] $0\leq 
\lambda^2\alpha^2<1-\frac{64Q^6}{27M^4}\left(1+\alpha^2\lambda^2\right)^4$
$\quad\quad$ \hfill Black hole region

\item[(ii)] $\lambda^2\alpha^2=1-\frac{64Q^6}{27M^4}
\left(1+\alpha^2\lambda^2\right)^4$
$\quad$ \hfill Extreme black hole line

\item[(iii)] $1-\frac{64Q^6}{27M^4}\left(1+\alpha^2\lambda^2\right)^4
<\lambda^2\alpha^2<1$ $\quad$ \hfill Naked singularity region

\item[(iv)] $\lambda^2\alpha^2=1$ $\quad\quad$ \hfill Null singularity line

\item[(v)] $1<\lambda^2\alpha^2\leq2$ $\quad\quad$ \hfill 
Pathological black hole region 

\hfill(black holes with closed timelike curves)

\end{description}
For $J>\sqrt{\frac{9M}{3\alpha}}$, the specific angular momentum 
parameter $\lambda$ turns complex, and the metric is ill-defined. 
The extreme case is given when $Q$ is connected to $\lambda$ 
through the relation $Q^6=\frac{27}{64}\frac{1-
\lambda^2\alpha^2}{1+\frac12\lambda^2\alpha^2}$. For the Penrose 
diagrams see \cite{lemoszanchin96,lemoszanchin96b}.

In the uncharged case, 
one has three regions only \cite{lemos1}, 
\begin{description}
\item[(i)] $0\leq\lambda^2\alpha^2<1$ \hfill Black hole region

\item[(ii)] $\lambda^2\alpha^2=1$\hfill Extreme black hole line

\item[(iii)] $1<\lambda^2\alpha^2\leq2$ \hfill Naked singularity region

\end{description}
For the Penrose diagrams of the uncharged case, see \cite{lemos2}.

The black hole as well as the naked singularity solutions can appear
through gravitational collapse
\cite{lemoscollapse1,lemoscollapse2,ghosh2000} and the non-rotating
configurations have thermodynamical properties related to
Reissner-Nordstr\"om black holes \cite{pecalemos1,pecalemos2,brill,vanzo}.

\vskip 0.5cm

\noindent
{\bf 4. The integrability conditions and Killing spinors 
for the charged configurations}
 
\vskip 3mm

In order to find the full covariant derivative we need to compute 
the spin connections for the tetrad (\ref{tetrads1}) using equation 
(\ref{spinconnection1}) with the spinor part set to zero (or,  
alternatively, solving the first Cartan equation 
$d e^a+{\omega^a}_b\wedge e^b=0$).
 
The non-zero spin connections are:
\begin{equation}
{\omega_t}^{01}=\frac12 \frac{1}{\alpha^2 r^2}\left(\Delta'-2\frac{\Delta}{r} 
                                                            \right) \, ,
                              \label{omegat01}
\end{equation}
\begin{equation}
{\omega_\phi}^{01}=-\frac12 \frac{\lambda}{\alpha^2 r^2} 
                    \left(\Delta'-2\frac{\Delta}{r}  \right) \, ,
                              \label{omegaphi01}
\end{equation}
\begin{equation}
{\omega_z}^{12}=-\frac12 \frac{\sqrt \Delta}{r} \, ,
                              \label{omegaz12}
\end{equation}
\begin{equation}
{\omega_t}^{13}=-\alpha \lambda \frac{\Delta}{r} \, ,
                              \label{omegat13}
\end{equation}
\begin{equation}
{\omega_\phi}^{13}= \frac{\sqrt\Delta}{\alpha r} \, 
                              \label{omegaphi13}
\end{equation}
with $'\equiv d\,/dr$. Then the supercovariant derivatives are
\begin{eqnarray}
\hat{\nabla}_t &=& \partial_t -i\alpha \left(-\frac{2Q_E}{\alpha r}+ 
2Q_M\lambda \alpha^2 z\right) \nonumber \\
&&
+ \left(\frac12 \alpha+\frac{i}{4}F_{ab}\gamma^{ab}\right)
\left(\frac{\sqrt \Delta}{\alpha r} \gamma_0 + 
r \lambda \alpha^2 \gamma_3\right)  \nonumber\\
&&
+\frac14\frac{1}{\alpha^2 r^2}\left(\Delta'-
2\frac{\Delta}{r} \right)\gamma_0\gamma_1
-\frac12 \alpha \lambda \frac{\Delta}{r} \gamma_1\gamma_3 \, , 
                                                        \label{nablat} \\
\hat{\nabla}_r &=& \partial_r + 
\frac12\frac{\alpha^2 r}{\sqrt \Delta}
\gamma_1 +  \frac{\alpha r}{\sqrt \Delta} 
\frac{i}{4}F_{ab}\gamma^{ab}\gamma_1 \, ,
                                                    \label{nablar}\\
\hat{\nabla}_z &=& \partial_z + 
\frac12 \alpha^2 r \gamma_2 
+ \alpha r \frac{i}{4}F_{ab}\gamma^{ab} \gamma_2 - 
\frac14  \frac{\sqrt \Delta}{r} \gamma_1\gamma_2 \, ,
                                                    \label{nablaz}\\
\hat{\nabla}_\phi &=& \partial_\phi -i\alpha 
\left( \frac{2Q_E\lambda}{\alpha r}
-2Q_M z \right)  \nonumber \\
&&
-\left(\frac12 \alpha+\frac{i}{4}F_{ab}\gamma^{ab}\right)
\left(\frac{\lambda\sqrt{\Delta}}{\alpha r}\gamma_0 + r\gamma_3\right) 
\nonumber\\
&&
-\frac14 \frac{\lambda}{\alpha^2 r^2} \left(\Delta'-2\frac{\Delta}{r}\right)
\gamma_0\gamma_1 + 
\frac12 \frac{\sqrt \Delta}{\alpha r} \gamma_1\gamma_3
\, ,
                                                    \label{nablaphi}
\end{eqnarray}
with $(i/4) F_{ab}\gamma^{ab}=(i/\alpha r^2)
(Q_E\gamma_0\gamma_1+Q_M\gamma_2\gamma_3)$.

Like in the case of Reissner-Nordstr\"om-anti de Sitter
\cite{romans92}, of Kerr-Newman-anti-de Sitter
\cite{caldarelliklemm98},  and for the other class of rotating
topological black holes in anti-de Sitter spacetimes
\cite{caldarelliklemm98} one finds that the super-Riemann tensor
(\ref{integrabilityconditions}) can be written as a product
\begin{equation}
\hat{R}_{mn} = {\cal P} {\cal G}_{mn} {\cal O} \,
                      \label{superriemann2}
\end{equation}
where $\cal P$ is a projection operator given by 
\begin{equation}
{\cal P} = \frac{\alpha r^2}{2Q} i F_{ab}\gamma^{ab}\gamma_1
= \frac{\alpha r^2}{2Q} i (-F_{01}\gamma_0 + F_{23}\gamma_1
\gamma_2 \gamma_3)  \, ,
                      \label{projectionoperator}
\end{equation}
${\cal G}_{mn}$ is $\gamma_{mn}$ times some function of $r$, and $\cal O$ 
is given by
\begin{equation}
{\cal O} = \frac{\sqrt\Delta}{\alpha r} +\alpha r \gamma_1 +
\sigma
{\cal P} \, ,
                      \label{operatorO1}
\end{equation}
where $\sigma\equiv \left[\frac{b}{4Q}-\frac{2Q}{\alpha r} 
(1-\tilde{\lambda}^2)\right]$, and we have put $\tilde{\lambda}\equiv
\lambda\alpha$, to simplify the notation.
In the real representation of the gamma matrices we find
\begin{equation}
{\cal O}=\pmatrix{ 
\frac{\sqrt\Delta}{\alpha r}-\alpha r & 0 & 
i\frac{Q_M}{Q}\sigma
& -i\frac{Q_E}{Q}\sigma
\cr 
0 & \frac{\sqrt\Delta}{\alpha r}+\alpha r & i\frac{Q_E}{Q}\sigma 
&  i\frac{Q_M}{Q}\sigma 
\cr
-i\frac{Q_M}{Q}
\sigma
& -i\frac{Q_E}{Q}
\sigma 
& \frac{\sqrt\Delta}{\alpha r}-\alpha r  & 0
\cr
i\frac{Q_E}{Q} \sigma
& -i\frac{Q_M}{Q}\sigma
& 0
&  \frac{\sqrt\Delta}{\alpha r}+\alpha r}\, .
                        \label{operatorO2}
\end{equation}
For $Q\neq0$, ${\cal G}_{mn}$ and $\cal P$ are non-singular
and the integrability conditions for the existence of Killing spinors 
are equivalent to the vanishing of the determinant of $\cal O$.
One then finds
\begin{eqnarray}
&{\rm det} \, {\cal O} =
\left[\left(\frac{\Delta}{\alpha^2r^2} -\alpha^2r^2\right) -
\sigma^2\right]^2
-4\frac{{Q_M}^2}{Q^2}\sigma^2\alpha^2 r^2 = \nonumber\\
&
\left[\frac{4Q^2}{\alpha^2r^2}\tilde{\lambda}^2(1-\tilde{\lambda}^2)-
\frac{b}{\alpha r}\tilde{\lambda}^2
-\frac{b^2}{4Q^2}\right]^2-
\left[\frac{2Q_Mb\alpha r}{4Q^2} - 4Q_M(1-\tilde{\lambda}^2)\right]^2
                       \label{determinant1}
\end{eqnarray}

Since  $Q\neq0$ the constraint implies either $\tilde{\lambda}=0$ (i.e.,
$J=0$), the non-rotating case, or $\tilde{\lambda}=1$, the null
singularity rotating case,  ($\tilde{\lambda}$ can also take negative 
values, in this case $\tilde{\lambda}=-1$, but the results do not 
differ, and we stick to the positive value of $\tilde{\lambda}$ always).
All these cases yield charged naked
singularities, which we analyze now.  
\begin{description} 
\item[(i)]
{\bf $\tilde{\lambda}=0$:} For no-rotation the determinant 
(\ref{determinant1}) simplifies to
\begin{eqnarray}
{\rm det} \, {\cal O} = \frac{1}{(4Q)^4} & 
\left[b^2-\left(2Q_Mb\alpha r-16 Q_MQ^2\right)\right]
\times \nonumber \\
&
\left[b^2+\left(2Q_Mb\alpha r-16 Q_MQ^2\right)\right] .
                       \label{determinant2}
\end{eqnarray}
Then ${\rm det}\,{\cal O}=0$ implies
\begin{equation}
Q_Mb\alpha=0 \, ,
                        \label{constraint}
\end{equation}
\begin{equation}
b^2=\pm 16 Q_MQ^2 \, .
                        \label{bogomolnyi}
\end{equation}
As stated in \cite{alonsoalercaetal00}, equation (\ref{constraint}) is
a constraint, while (\ref{bogomolnyi}) is the saturated Bogomol'nyi
bound for the configurations.  For $\alpha\neq0$ (which we 
usually assume, see \cite{gibbons3,todd,caldarelliklemm98} 
for $\alpha=0$), one finds $Q_M=0$
$\Rightarrow$ $b=0$, i.e., $M=0$.  Then $\Delta=4Q_E^2+\alpha^4r^4$.
This is a massless charged naked singularity with no-rotation, in the
nomenclature of Romans \cite{romans92} called a cosmic electric
monopole. Caldarelli and Klemm \cite{caldarelliklemm98} found this
supersymmetric solution and have displayed the corresponding Killing
spinors. Note, however, that when treating the toroidal case 
one has to change 
$b$, $Q_M$, and $Q$ defined for the cylindrical model as charges per 
unit length, to the bar quantities $\bar b$, $\bar Q_M$, and $\bar Q$, 
defined previously for the toroidal model. Then, since (\ref{bogomolnyi})
is not homogeneous in those three charges, it turns into 
$\bar{b}^2=\pm 16 \alpha \bar{Q}_M\bar{Q}^2$, where a new factor $\alpha$ has 
appeared, which is the integrability condition found in 
\cite{caldarelliklemm98} (see also \cite{alonsoalercaetal00}). In 
\cite{caldarelliklemm98} was then found another supersymmetric configuration
with $\alpha=0$, and $\bar{M}=0$.

\item[(ii)] {\bf $\tilde{\lambda}=1$:} The charged configuration is 
a rotating null singularity. The determinant (\ref{determinant1})
is now
\begin{equation}
{\rm det} \, {\cal O} =
\left[\frac{b}{\alpha r}\tilde{\lambda}^2+\frac{b^2}{4Q^2}\right]^2-
\left[\frac{2Q_Mb\alpha r}{4Q^2} - 4Q_M(1-\tilde{\lambda}^2)\right]^2\, .
                       \label{determinant3}
\end{equation}
However, since $\tilde{\lambda}=1$, then automatically $b=0$, and
(\ref{determinant3}) is trivially satisfied. In this case
$\Delta=\alpha^2r^2$ and the integrability condition ${\cal O}
\epsilon=0$ simplifies to
\begin{equation}
{\cal O} \epsilon= (2\alpha r) 
\left[\frac12 (1+\gamma_1)\right] \epsilon =0
\, , 
                       \label{Oepsilon}
\end{equation}
where $\frac12 (1+\gamma_1)$ is a projection operator. 
To find whether the solution is supersymmetric or not 
one should construct the Killing spinors. Using equations 
(\ref{nablat})-(\ref{nablaphi}) and (\ref{Oepsilon}) we find
\begin{eqnarray}
&\partial_t \epsilon + \left[i\frac{Q_E}{r}
\left(1+\gamma_0\gamma_3\right) -i 2Q_M\alpha^2z +
 i\frac{Q_M}{r}
\left(1-\gamma_0\gamma_3\right)\right]\epsilon =0
                                 \label{nablat2} \\&
\partial_r \epsilon + 
\left[ \frac{1}{2r}\gamma_1
+ \frac{i}{\alpha^2r^3}
\left(Q_E\gamma_0-Q_M\gamma_1\gamma_2\gamma_2\right)
 \right]\epsilon =0\, ,
                                            \label{nablar2}\\&
\partial_z \epsilon+ 
\frac{i}{r}
\left(Q_E\gamma_0\gamma_2-Q_M\gamma_3\right)\epsilon=0\, ,
                                                    \label{nablaz2}\\&
\partial_\phi \epsilon -
\left[i\frac{Q_E}{r}
\left(1+\gamma_0\gamma_3\right) -i 2Q_M\alpha^2z +
 i\frac{Q_M}{r}
\left(1-\gamma_0\gamma_3\right)\right]\epsilon =0 \, 
                                                    \label{nablaphi2}&
\end{eqnarray}
\end{description}
Using the techniques of Romans \cite{romans92} (see also
\cite{caldarelliklemm98}) we find that there are no Killing spinor
solutions to equations (\ref{nablat2})-(\ref{nablaphi2}). 
Thus the only supersymmetric charged toroidal configurations are the static 
ones studied in \cite{caldarelliklemm98}. We now turn to the 
uncharged case.

\newpage
\vskip 0.5cm

\noindent
{\bf 5. The Killing spinors for the extreme rotating uncharged 
black hole spacetime }
\vskip 3mm

When $Q=0$ (i.e., $Q_E=Q_M=0$) the integrability conditions 
(\ref{integrabilityconditions}) yield
\begin{equation}
\hat{R}_{mn}={\cal G}_{mn}b=0 \, ,
                               \label{integrabilitycondition2}
\end{equation}
where ${\cal G}_{mn}$ is some function of $r$ times $\gamma_m\gamma_n$.
Thus $b=0$ which, since 
$b=4M\frac{1-\tilde{\lambda}^2}{1+\frac12\tilde{\lambda}^2}$, 
yields 2 cases: (i) $M=0$, the anti-de Sitter space 
with cylindrical or toroidal topology, and (ii) $\tilde{\lambda}=1$, 
the extreme rotating toroidal black hole \cite{lemos1}. We now analyze 
each case. 
\begin{description}
\item[(i)] {\bf $M=0$:} The Killing spinor equations 
(\ref{nablat})-(\ref{nablaphi}) reduce to
\begin{eqnarray}
\hat{\nabla}_t \epsilon&=& \left[\partial_t  + 
\frac12 \alpha^2 r 
\left(\gamma_0+\gamma_0\gamma_1\right)\
\right]\epsilon=0, , 
                                                        \label{nablat3} \\
\hat{\nabla}_r\epsilon &=& \left[
\partial_r + \frac{1}{2r}\gamma_1 \right]\epsilon=0\, ,
                                                    \label{nablar3}\\
\hat{\nabla}_z\epsilon &=& \left[\partial_z + \frac12 \alpha^2 
r\left(\gamma_2-\gamma_1\gamma_2 \right) \right]\epsilon=0\, ,
                                                    \label{nablaz3}\\
\hat{\nabla}_\phi \epsilon&=& \left[
\partial_\phi - \frac12 \alpha r 
\left( 
 \gamma_3 -\gamma_1\gamma_3 \right)\right]\epsilon=0
 \, .
                                                    \label{nablaphi3}
\end{eqnarray}
Solving this set of equations yields
\begin{eqnarray}
\epsilon &=& \left[\sqrt{r}\frac12(1-\gamma_1)+ \frac{1}{\sqrt{r}}\frac12
(1+\gamma_1)\right]
\left[1-\gamma_0(1+\gamma_1)\frac{t}{2}\right] \nonumber\\
&&
\left[1-\gamma_2(1+\gamma_1)\frac{z}{2}\right]
\left[1+\gamma_3(1+\gamma_1)\frac{\phi}{2}\right]\epsilon_0 \, ,
                                                    \label{spinor3a}
\end{eqnarray}
where $\epsilon_0$ is a constant Dirac spinor
\begin{equation}
\epsilon_0=\pmatrix{ a \cr b \cr c \cr d}\, .
                        \label{spinorconstant3}
\end{equation}
In order that $\epsilon_0$ is independent of the angle $\phi$ 
one has $\frac12 (1+\gamma_1)\epsilon_0=0$, which amounts to 
say that $b=d=0$. Thus
\begin{equation}
\epsilon= \sqrt{r} \left[\frac12 (1-\gamma_1)\right] \epsilon_0\, ,
                        \label{spinor3b}
\end{equation}
defining two linearly independent Dirac spinors (see also
\cite{caldarelliklemm98}). 
\item[(ii)] {\bf $\tilde{\lambda}=1$:} This is the extreme black hole 
\cite{lemos1,lemoszanchin96}. 
Using the integrability conditions (i.e., ${\cal O} \epsilon=0$) 
in equations (\ref{nablat})-(\ref{nablaphi}) one gets
\begin{eqnarray}
\hat{\nabla}_t \epsilon &=& \left[\partial_t  + \frac12 \alpha^2 r 
\left(\gamma_0+\gamma_3+\gamma_0\gamma_1-\gamma_1\gamma_3\right)
\right]\epsilon=0\, , 
                                                        \label{nablat4} \\
\hat{\nabla}_r \epsilon  &=& \left[ \partial_r + \frac{1}{2r}\gamma_1
\right]\epsilon=0 \, ,
                                                    \label{nablar4}\\
\hat{\nabla}_z  \epsilon &=& \left[\partial_z + \frac12 \alpha^2 
r\left(\gamma_2-\gamma_1\gamma_2 \right)
\right]\epsilon=0 \, ,
                                                    \label{nablaz4}\\
\hat{\nabla}_\phi &=& \left[\partial_\phi - \frac12 \alpha r \left( 
\gamma_0 + \gamma_3 + \gamma_0\gamma_1-\gamma_1\gamma_3 \right)
\right]\epsilon=0
 \, .
                                                    \label{nablaphi4}
\end{eqnarray}

Solving this set of equations yields
\begin{eqnarray}
\epsilon &=& \left[\sqrt{r}\frac12(1-\gamma_1)+ \frac{1}{\sqrt{r}}\frac12
(1+\gamma_1)\right]
\left[1-(\gamma_0+\gamma_3)(1+\gamma_1)\frac{t}{2}\right] \nonumber\\
&&
\left[1-\gamma_2(1+\gamma_1)\frac{z}{2}\right]
\left[1+(\gamma_0+\gamma_3)(1+\gamma_1)\frac{\phi}{2}\right]\epsilon_0 \, ,
                                                    \label{spinor4a}
\end{eqnarray}
where, again, $\epsilon_0$ is a constant Dirac spinor
\begin{equation}
\epsilon_0=\pmatrix{ a \cr b \cr c \cr d}\, .
                        \label{spinorconstant4}
\end{equation}
In order that $\epsilon_0$ is independent of the angle $\phi$ 
one has $\frac12 (1+\gamma_1)\epsilon_0=0$, which amounts to 
say that $b=d=0$. Thus, 
\begin{equation}
\epsilon= \sqrt{r} \left[\frac12 (1-\gamma_1)\right] \epsilon_0\, ,
                        \label{spinor4b}
\end{equation}
defining, again, two linearly independent Dirac spinors. The extreme 
uncharged black hole has two supersymmetries,  the same number as 
the background spacetime, and in these coordinates the Killing  
spinors have the same expression. 

It is important to show that the extreme black hole has zero temperature. 
Indeed \cite{lemos1}, 
\begin{equation}
T={\alpha\over2\pi} {3\over2} M^{1/3}
\left[\frac{1-\tilde{\lambda}^2}{2(1+\frac12\tilde{\lambda}^2)}\right]^{1/3}
\left(1-\tilde{\lambda}^2\right)^{1/2}\, .
                        \label{temperature}
\end{equation}
For the extreme supersymmetric black hole, $\tilde{\lambda}=1$, 
yielding a black hole of zero temperature.

\end{description}

\vskip 0.5cm

\noindent
{\bf 6. Conclusions}
 
\vskip 3mm

We found that the uncharged toroidal (or cylindrical) black hole must
rotate extremally to be supersymmetric. There are no breaks of
supersymmetry: there are two supersymmetries, as in the original
spacetime.  This black hole has zero temperature.  The uncharged black
hole with planar topology cannot be supersymmetric since it does not
rotate. In this sense, the topology of the spacetime plays a
role in the supersymmetry analysis. 

For charged configurations we found that the supersymmetric ones are
static naked singularities with zero magnetic charge.  There are no
rotating supersymmetric charged configurations.

\vskip 1.cm

\section*{Acknowledgments} 
I thank Observat\'orio Nacional-MCT do Rio de Janeiro for hospitality
while this work was being initiated. I thank the Theory Division -
CERN, where the most part of this work has been done, for hospitality and for
providing an Associate grant for three months. I also thank Vilson
Zanchin for conversations, and FAPERGS for a grant to stay at
Universidade Federal de Santa Maria, Rio Grande do Sul, where this
work was finished.


\end{document}